\documentclass[superscriptaddress,prl,twocolumn]{revtex4}

\usepackage{amstext,amsmath,amssymb,amsfonts,bbm}
\usepackage{euscript}

\def\be{\begin{equation}}
\def\ee{\end{equation}}
\def\bes{\begin{eqnarray}}
\def\ees{\end{eqnarray}}
\def\bay{\begin{array}}
\def\ear{\end{array}}

\def\etal{\textit{et al.}}

\def\1{{{\mathbbm 1}}}
\def\bnab{{\mbox{\boldmath{$\nabla$}}}}
\def\anab{{\mbox{\sf\boldmath{{$\nabla$}}}}}
\def\half{\mbox{$1\over2$}}

\def\arb{\mathsf{b}}
\def\are{\mathsf{e}}
\def\af{{\sf f}}
\def\ak{{\sf k}}
\def\au{\mathsf{u}}

\def\sg{\mbox{\sl g}}

\def\bE{\mathbf{E}}

\def\bob{\mathbf{b}}
\def\boe{\mathbf{e}}
\def\bof{\mathbf{f}}

\def\bk{\mathbf{k}}

\def\hbb{{\hat{\bob}}}
\def\hbe{{\hat{\boe}}}
\def\hbf{{\hat{\bof}}}
\def\hbk{{\hat{\bk}}}
\def\hbz{\hat{\mathbf{z}}}

\def\eR{\EuScript{R}}

\def\co{{\cal O}}

\newcommand{\Omegab}{\mbox{\boldmath$\Omega$}}
\newcommand{\omegab}{\mbox{\boldmath$\omega$}}

\def\htheta{{\hat{\theta}}}
\def\hphi{{\hat{\phi}}}

\def\mout{{\mathrm{out}}}
\def\mrin{{\mathrm{in}}}

\begin{document}
\title{Polarization rotation, reference frames and Mach's principle}
\author{Aharon Brodutch}
\affiliation{Department of Physics \& Astronomy, Macquarie University, Sydney NSW 2109, Australia}
\author{Daniel R. Terno}
\affiliation{Department of Physics \& Astronomy, Macquarie University, Sydney NSW 2109, Australia}
\affiliation{Centre for Quantum Technologies, National University of Singapore, Singapore 117543}
\begin{abstract}
 Polarization of light rotates in a gravitational field. The accrued phase is operationally meaningful only
with respect to a local polarization basis. In stationary space-times, we construct local reference frames
that allow us to isolate the Machian gravimagnetic effect from the geodetic (mass) contribution to the
rotation. The Machian effect is supplemented by the geometric term that arises from the choice of standard
polarizations. The phase accrued along a close trajectory is gauge-independent and is zero in the
Schwarzschild space-time. The geometric term may give a dominant contribution to the phase. We
calculate polarization rotation for several trajectories and find it to be more significant than is usually
believed, pointing to its possible role as a future gravity probe.

\end{abstract}

\maketitle

\textit{Introduction.} Description of electromagnetic   waves  in terms of rays and  polarization vectors is
 a theoretical basis for much of optics \cite{bw} and observational astrophysics \cite{pad1}.
  Light rays in  general relativity (GR) are null geodesics. Their   tangent four-vectors $\ak$, $\ak^2=0$, are orthogonal to the spacelike polarization vectors $\af$, $\af^2=1$. Polarization is
 parallel-transported along the rays \cite{mtw,chandra},
\be
\anab_\ak\ak=0, \qquad \bnab_\ak\af=0, \label{ray}
\ee
where $\anab_\ak$ is a covariant derivative along $\ak$. These equations were solved for a variety of backgrounds (\cite{chandra,skrot,gravimag,fayos, kerr-farpol, lense, burns, kopeikin, faraoni08,gab11}), ranging from the  weak field limit of a single  massive body   to gravitational lenses  and  gravitational waves.

Solutions of \eqref{ray} predict  polarization  rotation in a gravitational field. Mass of the gravitating system is thought to have only a trivial effect, while its spin and higher moments cause the  gravimagnetic/Faraday/Rytov-Skrotski\u{\i} rotation (phase) \cite{skrot,gravimag,fayos, kerr-farpol}. A version of Mach's principle \cite{mach}, interpreted as presence of the Coriolis acceleration due to frame dragging around a rotating massive body \cite{mtw,chandra}, was first proposed to calculate polarization rotation in \cite{god}. It provides a convenient conceptual framework for the gravimagnetic effect \cite{gravimag}.

Polarization  is operationally meaningful only if its direction is compared with some standard polarization basis (two linear polarizations, right- and left-circular polarizations, etc.).  Insufficient attention to the definition of the standard polarizations is  one of the reasons for  the disparate values of the rotation angle $\Delta\chi$  that are found in the literature (see \cite{bt-d} for the discussion). In this paper we investigate the role of local reference frames in establishing $\Delta\chi$, focusing primarily on  stationary space-times.

We derive the  equation for polarization rotation for an arbitrary  choice of the polarization basis. In addition to  the expected Machian term \cite{gravimag,god} this equation contains  a reference-frame term, which makes a dominant contribution to the net rotation in the examples we consider. Only a particular choice of the standard polarizations allows stating that the  mass of the gravitating body does not lead to a phase along an open orbit, and we demonstrate how this gauge can be constructed by local observers. On the other hand, on closed trajectories the reference-frame term results in a gauge-independent contribution.

Taking  Kerr space-times as an example we calculate $\Delta\chi$ for several trajectories, and find the phase to be greater than it was commonly believed.  As a result, polarization rotation  may become a basis for  alternative precision tests of GR.

 We use $-+++$ signature, set $G=c=1$ and use Einstein summation convention in all dimensions. Three-dimensional  vectors are written in boldface and the unit vectors are distinguished  by carets, such as $\hbe$.

\textit{Trajectories and rotations}. To every local observer with a four-velocity $\au$ we attach an orthonormal tetrad, with the time axis defined by $\are_{0}\equiv\au$ \cite{mtw}.
In stationary space-times  a tetrad of a static observer is naturally related to the Landau-Lifshitz 1+3 formalism \cite{ll2}.
Static observers follow the congruence of timelike Killing vectors that  defines a projection from the  space-time manifold $\cal{M}$ onto a three-dimensional space $\Sigma_3$, $\pi:\mathcal{M}\rightarrow \Sigma_3$.

Projecting is performed in practice by dropping the timelike coordinate of an event. Vectors are projected by a push-forward map
$\pi_*\ak=\bk$ in the same way. For a static observer the three spatial basis vectors of the local orthonormal tetrad are projected  into an orthonormal triad, $\pi_*\are_{m}= \hbe_{m}$, $\hbe_m\!\cdot\!\hbe_n=\delta_{mn}$, $m,n=1,2,3$.

The metric $\sg$ on $\mathcal{M}$ can be written in terms of a three-dimensional scalar $h$, a vector $\mathbf{g}$, and a three-dimensional metric $\gamma$ on $\Sigma_3$ \cite{ll2} as
\be
ds^2=-h(dx^0-\sg_mdx^m)^2+dl^2,
\ee
where $h=-\sg_{00}$, $\sg_{m}=-\sg_{0m}/\sg_{00}$ and the  three-dimensional distance is  $dl^2=\gamma_{mn}dx^mdx^n$. 
 The inner product of three-vectors will always refer to this metric. 
 Vector products and differential operators are defined as appropriate dual vectors \cite{ll2}. 

For example,  the Kerr metric in the Boyer-Lyndquist coordinates \cite{mtw,chandra,ll2} leads to
\be
h=1-\frac{2M r}{\rho^2}, \qquad \sg_{0i}=-\delta_{i\phi}\frac{2Ma r}{\rho^2}\sin^2\theta,
\ee
 where $M$ is the mass of a gravitating body, $a=J/M$ is the angular momentum per unit mass, $\rho^2=r^2+a^2\cos^2\theta$ and $\delta_{ij}$ is Kronecker's delta.

Using the relationship between four- and three-dimensional covariant derivatives, $\anab_\mu$ and $D_m$, respectively, the propagation equations \eqref{ray} in a stationary space-time \cite{fayos, gravimag} result in a joint rotation of unit polarization and tangent vectors  \cite{bt-d},
\be
\frac{D\hbk}{d\lambda}=\Omegab\times\hbk, \qquad
\frac{D\hbf}{d\lambda}=\Omegab\times\hbf, \label{3dprot}
\ee
where $\lambda$ is the affine parameter and $\hbk=\bk/k$. The angular velocity of rotation $\Omegab$ is given by
\be
\Omegab=2\omegab-(\omegab\!\cdot\!\hbk)\hbk-\bE_g\times\bk,
\ee
where
\be
\omegab=-\half k_0\mathrm{curl}\ \!\mathbf{g}, \qquad \bE_g=-\frac{\nabla h}{2 h}.
\ee

\textit{Reference frames.}  For each momentum direction $\hbk$ polarization is specified with respect to two standard polarization vectors $\hbb_{1,2}$ . In a flat space-time these are  uniquely fixed by Wigner's little group construction \cite{wig,lpt1}. The standard reference momentum is directed along  an arbitrarily defined $z$-axis, with $x$ and $y$ axes defining the  two linear polarization vectors. A direction  $\hbk$  is determined by the spherical angles $(\theta,\phi)$.  The standard rotation that brings the $z$-axis to $\hbk$ is defined \cite{wig} as a rotation around the $y$-axis by $R_y(\theta)$ that is followed by the rotation $R_z(\phi)$ around the $z$-axis, so  $R(\hbk)\equiv R_z(\phi)R_y(\theta)$. Standard polarizations  for the direction $\hbk$ are defined as $\hbb_1\equiv R(\hbk)\hat{\mathbf{x}}$ and  $\hbb_2\equiv R(\hbk)\hat{\mathbf{y}}$. An additional gauge fixing promotes $\hbb_{i}$ to  four-vectors, so a general linear polarization vector is written as
\be
\af=\cos\chi\arb_1(\ak)+\sin\chi\arb_2(\ak). \label{poldec}
\ee
 We adapt a gauge  in which polarization  is orthogonal to the observer's four-velocity, $\au\!\cdot\!\af=0$, so in a three-dimensional form the transversality condition reads as  $\bk\!\cdot\!\hbf=0$.

 The net polarization rotation  that results from a Lorentz transformation can be read off either by using the definition of the corresponding little group element or by referring the resulting polarization vector $\af'$ to the new standard polarization vectors $\arb_i(\ak')$ \cite{lpt1}. On a curved background the standard polarization triad $(\hbb_1, \hbb_2,\hbk)$ should be defined  at every point.

 Assume that two standard linear polarizations are selected.
  By setting $\hbf=\hbb_1$ at the starting point of the trajectory
we have $\sin\chi(\lambda)=\hbf(\lambda)\!\cdot\hbb_2(\lambda)$, so 
\be
\frac{d\chi}{d\lambda}=\frac{1}{\hbf\!\cdot\!\hbb_1}\left(\frac{D\hbf}{d\lambda}\!\cdot\! \hbb_2+\hbf\!\cdot\!\frac{D\hbb_2}{d\lambda}\right)=\omegab\!\cdot\!\hbk+\frac{1}{\hbf\!\cdot\!\hbb_2}\hbf
\!\cdot\!\frac{D\hbb_2}{d\lambda}. \label{defrot}
\ee
This is the desired phase evolution equation. The term $\omegab\!\cdot\!\hbk$ is the  Machian effect as it was postulated in \cite{god}. The second term --- the reference-frame contribution --- was missing from the previous analysis. However, in the examples below the reference-frame term dominates  the Machian effect by  one power of the relevant large parameter.

In general, a rigid rotation of momentum and polarization 
leads to a non-zero polarization rotation $\Delta\chi$ with respect to the standard directions \cite{wig, lpt1}. Right- and left-circular polarizations remain invariant, but acquire Wigner phase factors $e^{\pm i\Delta\chi}$, respectively.

The statement of zero accrued phase $\Delta\chi=0$  in the Schwarzschild space-time \cite{skrot, gravimag,kopeikin}, where $a=0$, so $\omegab=0$ and $\Omegab=-\bE_g\times\bk$, is correct only in a particular gauge.  \textit{Defining} this phase to be zero   allows us to make a  choice of standard polarizations that does not require references to a parallel transport or communication between the observers. This gauge construction  is based on the following property, which is proved at the end of this Letter.

\textit{Property 1}. A rotation $R_{\hbb_2}(\alpha)$  around the $\hbb_2$-axis, where the polarization triad  $(\hbb_1, \hbb_2,\hbk)$  is obtained from $(\hat{\mathbf{x}},\hat{\mathbf{y}},\hat{\mathbf{z}})$ by the standard rotation $R(\hbk)$, does not introduce a phase. \hfill $\blacksquare$

In the Schwarzschild space-time  we see that if the $z$-axis is oriented along the direction of the free-fall acceleration $\bE_g$,
  and   $\hbb_2\propto \bE_g\times\bk=\Omegab$,  no polarization phase is accrued. As a photon propagates along its planar trajectory, its momentum $\hbk$ and  linear polarization $\hbb_1$ are rotated around the  direction $\hbb_2$, which is perpendicular to the plane of motion.

   In a general stationary space-time we construct the polarization basis by defining a local $z$-axis along the free-fall acceleration, $\hbz\equiv\mathbf{w}/|\mathbf{w}|$, and define the standard polarization  direction $\hbb_2$ to be perpendicular to the momentum and the local $z$-axis,
    \be
    \hbb_2\equiv\mathbf{w}\times\hbk/|\mathbf{w}\times\hbk|.
    \ee
 This convention, that we will call the Newton gauge, is consistent: if we set $\hbz=-\hat{\mathbf{r}}$ in the flat space-time, then  no phase is accrued as a result of the propagation. In addition to being defined by local operations, the Newton gauge has two further advantages. First, it does not rely on a weak field approximation for definition of the reference direction. Second,  if the trajectory is closed or self-intersecting, the reference direction $\hbz$ is the same at the points of the intersection.

\textit{Examples.} We discuss polarization rotation in a Kerr space-time in two different settings, giving the calculational details in \cite{bt-d}.  To simplify the  expressions  we   define the affine parameter \cite{mtw,chandra}  by fixing the energy $E=-k_0=1$. Trajectories are  calculated using Carter's conserved quantity $\eta$ \cite{chandra,ll2}.

The light is emitted from the point $(r_\mrin,\theta_\mrin,0)$ and observed at $(r_\mout,\theta_\mout, \phi_\mout)$. To simplify the exposition we assume a distant observation point ($r_\mout\rightarrow \infty$). In one scenario we consider a trajectory in the outgoing principal null geodesic congruence \cite{chandra,fayos}. In a ``scattering" scenario the light is emitted and observed far from the gravitating body ($r_\mrin,r_\mout\gg M,a$).

Outgoing geodesics in principal null congruence have $k^m=(1,0,a/\Delta)$, $\Delta\equiv r^2-2Mr+a^2$, along the entire trajectory, and can be easily integrated \cite{chandra}. Taking into account that the Machian term  $\omegab\!\cdot\!\hbk\sim\co(r^{-3})$ in Eq~\eqref{defrot} is dominated by the reference-frame term, the leading order of polarization rotation is found by a direct integration
\be
\sin\Delta\chi=-\frac{a}{r_\mrin}\cos\theta_\mrin+\co(r_\mrin^{-2}), \label{maxef}
\ee
in agrement with \cite{fayos}.

 Evolution of polarization along a general trajectory is most easily  obtained by using Walker-Penrose conserved quantity $K_1+iK_2$ \cite{chandra}.
We adapt the calculation scheme of \cite{kerr-farpol}.
Trajectories  reach the asymptotic outgoing value of the azimuthal angle $\theta_\mout$, and the conservation laws  determine the  outgoing momentum as
\be
k^{\hat{\mu}}=(1,1,\pm\beta_\mout/r,D/r \sin\theta_\mout),
\ee
where $D=L_z/E$ is a scaled angular momentum  along the $z$-axis, and $\beta_\mout^2\equiv\eta+a^2\cos^2\theta_\mout-D^2\cot^2\theta_\mout$. Initial momentum in the scattering scenario satisfies a similar expression.

Using the transversality of polarization  and the gauge condition $f^{\hat{0}}=0$ we select $f^\htheta$ and $f^\hphi$ as two  independent components. 
Walker-Penrose constants are bilinear in  the components of polarization and momentum. Hence one obtains a transformation matrix $R$,
\be
\left(\begin{array}{c}
f^\htheta_\mout \\
f^\hphi_\mout
\end{array} \right)=R\left(\begin{array}{c}
f^\htheta_\mrin \\
f^\hphi_\mrin
\end{array} \right).
\ee
Polarization is expressed in the basis $\hbb_1$, $\hbb_2$. Matrices $N$ connect $(f^\htheta,f^\hphi)$ with $(f^1, f^2)$. As a result the evolution is represented by the matrix
\be
T=N_\mout R N_\mrin^{-1}.
\ee
If initial polarization is $\hbf=\hbb_1^\mrin$, then the rotation angle is given by $\sin\chi=T_{21}$ \cite{bt-d}.

We present a  special case of the scattering scenario  that allows an easy comparison with the literature \cite{bt-d} and highlights the importance of a proper treatment of  reference frames.   If the initial propagation direction is parallel to the $z$-axis with the impact parameter $s$,
then the polarization is rotated by
\be
\sin\chi=\frac{4Ma}{\Lambda^2}+\frac{15M^2a}{4\Lambda^3}+\co(\Lambda^{-4}),
\ee
where
\be
\Lambda^2\equiv D^2+\eta=s^2-a^2.
\ee
The antiparallel initial direction gives the opposite sign.

At the leading order in $s^{-1}$ we can assume $\hbf\approx\hbb_1$ and directly integrate  Eq.~\eqref{defrot}.  The result shows that the leading contribution to the polarization rotation comes from the reference-frame term, while  the Machian term $\omegab\!\cdot\!\hbk$ at this order   does not contribute to the integral. The latter result was used to justify the view that the polarization rotation is $1/r_\mathrm{min}^3$ effect, where $r_\mathrm{min}\sim s$ is the minimal distance from the gravitating body.

\textit{Gauge-invariant phase.} Consider a  basis of 1-forms $(\sigma^1, \sigma^2, \sigma^3)$ that is dual to  $(\hbb_1, \hbb_2,\hbk)$. A matrix of connection 1-forms $\omega$ is written with the help of Ricci rotation coefficients $\omega^a_{cb}$ as $\omega^a_{\ b}=\omega^a_{cb}\sigma^c$ \cite{mtw, fra}.
Taking into account Eq.~\eqref{poldec}, and the antisymmetry of the connections $\omega^a_{\ b}=\omega_{ab}=-\omega_{ba}$, we find
\be
\frac{D\hbf}{d\lambda}=(-\hbb_1 f^2+\hbb_2 f^1)\left(\frac{d\chi}{d\lambda}-\omega^1_{32}k\right)+
\bk(\omega^3_{31}f^1+\omega^3_{32}f^2).
\ee
A comparison with Eq. \eqref{3dprot} leads to the identification $\Omega^1=\Omegab\!\cdot\!\hbb_1=k\omega^3_{32}$, $\Omega^2=-k\omega^3_{31}$, where $k=|\bk|$,  and an alternative equation for the polarization rotation,
\be
\frac{d\chi}{d\lambda}=\omegab\!\cdot\!\hbk+\omega^1_{32}k.
\ee
Freedom of choosing the polarization frame $(\hbb_1,\hbb_2)$ at every point of the trajectory is represented by a SO(2) rotation $R_\hbk\big(\psi(\lambda)\big)$. Under its action the connection transforms as $\omega\rightarrow R\omega R^{-1}+R^{-1}dR$ \cite{fra}, so $d\chi/d\lambda\mapsto d\chi/d\lambda+ d\psi/d\lambda.$

  A closed photon trajectory (i.e., $\ak^\mrin=\ak^\mout$, $x^m_\mrin=x^m_\mout$) may occur ``naturally" (as a result of the initial conditions) or with judiciously positioned mirrors. The resulting gauge-invariant phase is
 \be
 \Delta\chi=\oint \omegab\!\cdot\!\hbk d\lambda +\oint \omega^1_{32}k d\lambda. \label{gauge-inv}
 \ee
Given a trajectory with a tangent vector $\bk$  one can define a SO(2)  line bundle with the connection $\bar{\omega}=\omega^1_{32}k$, similarly to the usual treatment of  geometric phase \cite{fra,geom}. Using Stokes' theorem the reference-frame term can be rewritten as a surface integral of the  bundle curvature $\bar{\theta}=d\bar{\omega}$, as $\int \bar{\omega}=\int\!\int \bar\theta$.
A more practical expression follows from our previous discussion: $
\Delta\chi=\arcsin\hbf_\mout\!\cdot\!\hbb_2$.

Conservation of $K_1$ and $K_2$  in the Kerr space-time ensures that if a trajectory is closed as a result of the initial conditions, then $\hbf_\mout=\hbf_\mrin$ and $\Delta\chi=0$. The Newton gauge is designed to give a zero phase along any trajectory in the Schwarzschild space-time. As a result of the gauge invariance of Eq. ~\eqref{gauge-inv}, no phase is accrued along a closed trajectory in the Schwarzschild space-time, regardless of the gauge convention.

As an example of a non-zero phase  along a closed trajectory consider the following combination of the scattering scenarios.   The trajectory  starts parallel to the axis of rotation with an impact parameter $s_1$ (and the initial angle $\theta_1=\pi$). Far from the gravitating body (so its influence can be ignored), the outcoming photon is twice reflected and sent in again with the impact parameter $s_2$ and the initial angle $\theta_2=0$. After the second scattering   and appropriate reflections it is returned to the initial position with the initial value of the momentum. Then
\be
\Delta\chi={4aM}\big(\Lambda_1^{-2}-\Lambda_2^{-2}\big)+\co(\Lambda^{-3}).
\ee

\textit{Outlook.}  Observation of  the frame-dragging  effects
is the last of the ``classical" tests of GR \cite{will} that has not yet been performed with a sufficient accuracy. In addition to Gravity Probe-B \cite{gp-b} and LAGEOS \cite{lageos} experiments, there are proposals to use Sagnac interferometry with ring-laser gyroscopes \cite{optgyro}, or  matter or quantum optical interferometry \cite{interf}. The main difficulty in these tests  is the necessity to isolate a much larger geodetic effect that is caused by the Earth's mass.

 Polarization phase (along a closed trajectory or in the Newton gauge for an open path) is insensitive to it.
We take Eq.~\eqref{maxef} to estimate the polarization rotation in the near-Earth environment. The Earth angular momentum is $J_\oplus=5.86\times 10^{40}$ cm$^2$g\,sec$^{-1}$ \cite{inertia}.
 We send the photon  on a geodetic from the principal null congruence starting at some $(r_\mrin, \theta_\mrin)$ and detect it at infinity (otherwise we have to correct by the term $a\sin\theta_\mrin/r_\mout$). Assuming $r_\mrin=12\,270$ km (the semimajor axis of the LAGEOS satellite orbit \cite{lageos}), we obtain $\Delta\chi\sim 55$\,arc\,msec in a single run. For comparison, the relevant precession rates for the Gravity Probe-B and LAGEOS experiments are 39\,arc\,msec/yr and 31\,arc\,msec/yr, respectively.

We showed that polarization rotation in a gravitational field is obtained from both the Machian Coriolis acceleration and a reference-frame term, and the latter may be  dominant.  It is responsible for a gauge-independent geometric phase that is accrued on a closed trajectory in stationary space-times. The effect is proportional to the angular momentum of a gravitating body and possibly may be used in future gravity probe experiments.

\noindent \textit{Proof of Property 1.}
 If the triad $(\hbb_1,\hbb_2,\hbk)$ is rigidly rotated by $\eR$, it typically results in an angle between, say, $\eR \hbb_2$ and $\hbb_2({\eR\hbk})$. This is the Wigner's phase $\chi$ of photons. It can be read-off  from the definition \cite{wig,lpt1},
\be
R_z(\chi)\equiv R^{-1}(\eR\hbk)\eR R(\hbk).\label{wr}
\ee
However, if $\eR=R_{\hbb_2}(\alpha)$, then using the decomposition
\be
R_{\hbb_2}(\alpha)=R(\hbk)R_y(\alpha)R^{-1}(\hbk),
\ee
so $\hbk'=R_{\hbb_2}(\alpha)\hbk=\hbk'(\theta+\alpha,\varphi)$, the Wigner rotation matrix is
\begin{align}
R_z(\chi)&=R^{-1}(\hbk')R_{\hbb_2}(\alpha) R(\hbk) \nonumber \\
&= R_y^{-1}(\theta+\alpha)R_y(\theta)R_y(\alpha)=\1.
\end{align}

\textbf{Acknowledgments.} We thank B.-L. Hu, A. Kempf, G. Milburn, N. Menicucci, K. Singh, C. Soo and V. Vedral for comments and discussions.


\end{document}